# Influence of carbon on intraband scattering in Mg(B$_{1-x}$C$_x$)$_2$

M. Matusiak[1,*], K. Rogacki[1], N.D. Zhigadlo[2], and J. Karpinski[2]

1. Institute of Low Temperature and Structure Research, Polish Academy of Sciences,

P.O. Box 1410, 50-950 Wrocław, Poland

2. Laboratory for Solid State Physics, ETH Zurich, 8093 Zurich, Switzerland

**Abstract**

We report data on the Hall coefficient ($R_H$) of the carbon substituted Mg(B$_{1-x}$C$_x$)$_2$ single crystals with $x$ in the range from 0 to 0.1. The temperature dependences of $R_H$ obtained for the substituted crystals differ systematically at low temperatures, but all of them converge to the value of $1.8 \times 10^{-10}$ m$^3$C$^{-1}$ at room temperature. The $R_H(T)$ data together with results of the thermoelectric power and electrical resistivity measurements are interpreted within a quasi-classical transport approach, where the presence of four different conducting sheets is considered. The main influence of the carbon substitution on the transport properties in the normal state is associated with enhanced scattering rates rather than modified concentration of charge carriers. Presumably the carbon substitution increases the electron-impurity scattering mainly in the $\pi$ band.

[*] Corresponding author. Tel.: +48-71-3435021; Fax: +48-71-3441029; e-mail: M.Matusiak@int.pan.wroc.pl

Despite of extensive studies of the electronic structure of $MgB_2$ that have been carried out after the discovery of superconductivity in this compound [1], some aspects of its transport properties remains a matter of debate. One of vigorously discussed issues is a role of defects, which can be systematically introduced into the structure of $MgB_2$ by irradiation [2,3] or partial chemical substitutions (often Al for Mg or C for B) [4,5,6,7]. While some consequences of the substitution or irradiation, such as modification of the upper critical field ($H_{c2}$) or changes in the width of superconducting gaps are rather indubitable, there is still no agreement on microscopic mechanisms leading to these effects. A possible reason for this ambiguity is the complex electronic structure of $MgB_2$ that is characterized simultaneously by the quasi two-dimensional $\sigma$ band and the three-dimensional $\pi$ band, where each band consists of two sheets of the Fermi surface [8]. All four bands take part in the electronic transport in the *ab*-plane, and some aspects of these bands may be dramatically different [9,10,11]. In this work we present data on transport coefficients of carbon doped series of $Mg(B_{1-x}C_x)_2$ single crystals, and analyze results using minimal constraints.

For the Hall effect measurements, we chose single crystals of $Mg(B_{1-x}C_x)_2$ with $x = 0$ (unsubstituted), 0.02, 0.06, and 0.1. The thermoelectric power at room temperature was also evaluated for these samples and, additionally, for one with $x = 0.05$. The resistivity measurements were performed on the crystals with $x = 0$ and $x = 0.06$. The Hall coefficient ($R_H$) was measured by a standard procedure in the magnetic field of 13 T. Samples were rotated by 180° and the current direction was reversed many times to exclude the influence of mismatching of the Hall contacts positions and of detrimental emf's. Methods to measure the resistivity ($\rho$) and thermoelectric power ($S$) were described in detail in Refs. [7] and [12], respectively.

The crystals were grown under high pressure using a cubic anvil press. The applied pressure and temperature conditions for the growth of MgB2 single crystals were determined in our earlier study of the Mg-B-N phase diagram [13]. A mixture of Mg, B, BN and C (in the case of C-substituted crystals) was ground and cold-pressed into a pellet. Then, the pellet was put into a BN container of 8 mm internal diameter and 8 mm length. Both unsubstituted and C-substituted crystals were grown in similar way. First, pressure was applied using a pyrophylite pressure transmitting cube as a medium. Then, the temperature was increased during 1 h, up to the maximum of 1900–1950 °C, kept for 30 min, and decreased over 1–2 h. Small single crystals were selected (typically with dimensions of 0.5 x 0.3 x 0.04 mm$^3$) to reduce any influence of crystal imperfections. Magnetic measurements have been performed with a noncommercial superconducting quantum interference device (SQUID) magnetometer. The temperature dependence of the dc magnetization in an external magnetic fields of 0.3 - 0.5 mT was recorded for both zero-field-cooled and field-cooled conditions.

Partial substitution of boron by carbon in $Mg(B_{1-x}C_x)_2$ has been widely studied since it was found to increase $H_{c2}$ and enhance critical currents, despite a modest decrease of $T_c$ [7, 14,15,16]. This behavior could reflect changes in charge carriers concentration and, moreover, in an interplay between various microscopic mechanisms that determine scattering and coupling within and between the $\sigma$ and $\pi$ bands. However, one can expect that in some particular phenomena one mechanism dominates others. A hint that this is the case of high temperature transport properties of $Mg(B_{1-x}C_x)_2$ comes from the electrical resistivity data. The linear dependence of the room temperature resistivity ($\rho^{300K}$) on $x$ for series of the $Mg(B_{1-x}C_x)_2$ single crystals is shown in the inset in Figure 1, where our resistivity data agree well with those reported by T. Masui and coworkers [14]. The methodical studies of $Mg(B_{1-x}C_x)_2$ single crystals reveal a dramatic increase in the residual resistivity with $x$ (where $\rho_0$ is found to be approximately proportional to *x*), while systematic changes in a shape of the temperature dependent part of resistivity are not present [14]. The linear $\rho_0(x)$ dependence accompanied by the practically unaltered temperature dependent part of resistivity might be in principle considered as a result of a coincidental compensation occurring between different conductivity bands, however we propose a simpler scenario.

The electronic transport in Mg(B$_{1-x}$C$_x$)$_2$ can be described within the relaxation time approximation with the electrical conductivity equals to: $\sigma_i = n_i e^2 \tau_i / m_i^*$, where *i* denotes an index of the band and sheet (namely $i = \pi 1, \pi 2, \sigma 1, \sigma 2$), $n_i$ – charge carrier concentration, *e* – elementary charge, $\tau_i$ – relaxation time, $m_i^*$ – effective mass. The total electrical conductivity aggregates contributions from all conductivity bands: $\sigma_{total} = \sum_i \sigma_i$. We also assume that at room temperature the different mechanisms of charge carriers scattering in Mg(B$_{1-x}$C$_x$)$_2$ are simply additive, i.e. they obey the Matthiessen's rule. Therefore, the electrical resistivity of each sheet of the Fermi surface is a sum of two contributions: one related to scattering on impurities ($\rho_0$) and a second related to the temperature dependent scattering on phonons ($\rho_{e-ph}$):

$$\rho_i = \frac{m_i^*}{n_i e^2 \tau_i^{imp}} + \frac{m_i^*}{n_i e^2 \tau_i^{e-ph}}. \tag{1}$$

Because the temperature dependent part of the resistivity is nearly independent of carbon content, we conclude that changes of $\tau_i^{imp}(x)$ dominates the entire $\rho(x)$ dependence. Otherwise, if changes in $n_i(x)$ or $\tau_i^{e-ph}(x)$ were significant, we should see a systematical variation in the slope $d\rho/dT$ versus *x* that seems to be absent [14]. Since at room temperature $\rho_{total}^{300K}(x) = (a_{total} + b_{total} x)$ we presume that: $1/\tau_i^{300K} = 1/\tau_i^{e-ph} + 1/\tau_i^{imp} = a_i + b_i x$, where $a_{total}$, $b_{total}$, $a_i$, and $b_i$ are constants.

This simple picture can be verified by results of the Hall effect measurements presented in Fig. 2. The in-plane Hall coefficient for the unsubstituted MgB$_2$ single crystal rises with falling temperature and $R_H^{50K}$ is about 30% larger than $R_H^{300K}$, what is very similar to previously reported data [14,17]. A gradual substitution of boron by carbon results in the systematic changes of the slope $dR_H/dT$. The value of Hall coefficient for Mg(B$_{1-x}$C$_x$)$_2$ with $x = 0.06$ practically does not depend on temperature, and for $x = 0.10$, $R_H^{50K}$ is even lower than $R_H^{300K}$. At this point we focus on $R_H^{300K}$ that one can find surprisingly independent of $x$. Carbon substitution introduces additional electrons to the electronic structure of MgB$_2$ and at first glance one may expect that the Hall coefficient will reflect these changes. However, as it was mentioned above, the magnesium diboride is a multi-band conductor and in such a case the total Hall coefficient is a sum of the Hall coefficients of individual bands weighted by square of their electrical conductivities [18]:

$$R_H = \frac{\sum_i R_{H,i}(\sigma_i)^2}{\sigma_{total}^2}. \quad (2)$$

If shapes of conductivity bands are not altered significantly by doping, as in our case, i.e. $m_i^*$ is constant and the Hall concentration is proportional by factor $c_i$ to $n_i$, then:

$$R_{H,i}\sigma_i^2 = \frac{c_i}{n_i e}\left(\frac{n_i e^2 \tau_i}{m_i^*}\right)^2 = C_i n_i \tau_i^2. \quad (3)$$

As concluded from the preceding analysis $\tau_i^{300K} = 1/(a_i + b_i x)$ and changes of the relaxation time with $x$ overwhelm effects related to variation of $n_i$. Thus:

$$R_H^{300K}(x) = const. = \frac{\sum_i \frac{C'_i}{(a_i + b_i x)^2}}{(\sigma_{total}^{300K})^2}, \quad (4)$$

where $C'_i = C_i n_i$, and:

$$\sum_i \frac{C'_i}{(a_i + b_i x)^2} \propto \frac{1}{(a_{total} + b_{total} x)^2}. \tag{5}$$

Eqn. 5 is satisfied only if relation between each pair of $a_i$ and $b_i$ on the left side of the equation is the same as relation between $a_{total}$ and $b_{total}$. On the other hand, the $C'_i$ parameters can vary with each band and one should try to overcome this difficulty. A simple method is to generalize the two $\pi$ and two $\sigma$ bands and assume that $\tau_{\pi 1} = \tau_{\pi 2} = \tau_\pi$ and $\tau_{\sigma 1} = \tau_{\sigma 2} = \tau_\sigma$ [19,20]. However, as shown by H. Yang and coworkers in Ref. [21], the values of all four relaxation times can differ considerably. Thus, we use a weaker constrain, where $\tau$'s in two sheets of a given band ($\pi$ or $\sigma$) are not equal, but proportionally related. In view of the preceding discussion they are also assumed to be inversely proportional to $x$, i.e.: $\tau_{\pi 1} \propto \tau_{\pi 2} \propto 1/(a_\pi + b_\pi x)$ and $\tau_{\sigma 1} \propto \tau_{\sigma 2} \propto 1/(a_\sigma + b_\sigma x)$. It indicates that Eqn. 5 can be only fulfilled, when carbon substitution modifies relaxation times primarily in one band, while second remains almost unaffected. This conclusion is consistent with results presented in other reports, however there is no agreement which band is influenced by the carbon substitution. Some reports indicate that it is the $\pi$ band [22,23,24], whereas others suggest the $\sigma$ band [25,26].

At this point we comment the $R_H(T)$ dependences for $T < 250$ K, where the Hall coefficient at given temperature is no longer independent of $x$ (see Fig. 2). We think that probable reason for this gradual aberrance from the above discussed picture is a partial violation of the Mathiessen's rule, which holds well for temperatures higher than 1/5 of the Debye temperature ($\Theta_D$), but significant deviations are expected at lower $T$ [27]. And in fact, since $\Theta_D$ for $MgB_2$ is about 900 K [28], this is the region in our experiment where $R_H(T)$ for samples with different $x$ begins to diverge substantially.

In the next step the room temperature thermoelectric power ($S^{300K}$) of our single crystals has been measured and the results are shown in Fig. 3. Measurements were performed in the in-plane ($S_{ab}$) and out-of-plane ($S_c$) configuration. In both cases, the dependence of thermopower on carbon content is linear with the exemption of the unsubstituted sample, where $S$ raises over the linear regressions for about 4 $\mu$V/K. Such aberrance for the pure crystal should probably be assigned to the phonon-drag thermopower ($S_{ph}$) that can be easily suppressed by crystalline defects, thus it is supposed to vanish quickly in the carbon doped samples. A maximum of $S_{ph}$ should appears at $T_{ph} \approx \Theta_D/5$ [29], and since the effective Debye temperature of MgB$_2$ is ~ 900 K [28] it gives $T_{ph}$ ~ 180 K. Thus we anticipate a significant manifestation of the phonon-drag thermopower in pure MgB$_2$ even at room temperature.

We deal with the thermopower data in a way analogous to that used in analysis of the Hall effect. Namely, we separate the total thermoelectric power for contributions from single conducting sheets weighted by electrical conductivity:

$$S = \frac{\sum_i S_i \sigma_i}{\sigma}. \qquad (6)$$

The thermopower in carbon doped MgB$_2$ is almost a linear function of temperature [12], what indicates its metallic character. Hence we utilize the well known Mott-Jones formula:

$$S_i = -\frac{\pi^2 k_B^2 T}{3|e|}\left(\frac{1}{\sigma_i}\frac{\partial \sigma_i}{\partial \varepsilon}\right)_{\varepsilon=E_F}, \qquad (7)$$

which leads to:

$$S_i \sigma_i = -\frac{\pi^2 k_B^2 |e|T}{3m_i^*}\left(n_i \frac{\partial \tau_i}{\partial \varepsilon} + \tau_i \frac{\partial n_i}{\partial \varepsilon}\right)_{\varepsilon=E_F}, \qquad (8)$$

where $n_i$ and $\tau_i$ depend on carbon content. Again we suppose that the main influence on the transport properties is caused by the variations of relaxation times. Then by using the experimental result $S^{300K}(x) = c + fx$, we obtain:

$$S^{300K}(x) = c + fx = \frac{\sum_i \left( g_i \left( \frac{\partial \tau_i(x)}{\partial \varepsilon} \right)_{\varepsilon=E_F} + h_i \tau_i(x) \right)}{\sigma_{total}^{300K}(x)}, \quad (9)$$

where $c, f, g_i, h_i$ are constants.

If we assume that the energy dependence of $\tau$ for $|\varepsilon - \varepsilon_F| < 2k_B T$ can be effectively fitted with a linear relation, then $(\partial \tau_i / \partial \varepsilon)_{\varepsilon=E_F} = const.$, and:

$$\frac{c + fx}{a_{total} + b_{total}x} = \sum_i \left( g'_i + \frac{h_i}{a_i + b_i x} \right) = \sum_i \left( \frac{h'_i + g''_i x}{a_i + b_i x} \right). \quad (10)$$

The Eqn. 10, which is analogous to the Eqn. 5, turns out to be satisfied only if changes in one conductivity band marginalize another. However, this still does not suggest whether the $\pi$ or $\sigma$ band is affected. We think that an indication comes from the $S_c(x)$ dependence, which is qualitatively the same as $S_{ab}(x)$ (see Fig. 3). It is known that due to quasi two dimensional nature of two $\sigma$-band sheets, they do not contribute considerably to the out-of-plane transport in the normal state [8]. The calculated values of the effective mass along $c$ axis for the $\sigma 1$ and $\sigma 2$ sheets are more than two order of magnitude higher than the corresponding values for the $\pi 1$ and $\pi 2$ sheets, whereas all four values of $m_i^*$ are comparable in the $ab$-plane [11]. It means that in the out-of-plane transport the electrical conductivities ($\sigma_i$), which in Eqn. 6 weight contributions of the thermopower from a given Fermi sheet ($S_i$), are more than hundred times higher for the $\pi$ band than for the $\sigma$ band. In our experiment the slope of the $S_{ab}(x)$ dependence is only about two times higher than the slope of $S_c(x)$ and it is reasonable to assume that similar behavior of $S_{ab}(x)$ and $S_c(x)$ is caused by the same mechanism. It encourages us to conclude that in both configurations we see the influence of the carbon substitution on the intraband scattering in the $\pi$ band. This seems to be contrary to the popular belief that the carbon substitution in the boron plane should mainly increase the scattering in the $\sigma$ band and perhaps leave the $\pi$ band unaffected. However, as explained in Ref. [30], in a case when the boron plane is substituted, the magnesium plane buckles and this can enhance scattering in the $\pi$ band. The enhanced scattering in the $\pi$ band due to the C substitution has been derived from the point-contact spectroscopy experiments [22,23] and from the upper critical field measurements [24].

In summary, our results show that high temperature behavior of the transport coefficients of the $Mg(B_{1-x}C_x)_2$ single crystals can be qualitatively understood within the framework of the multiband quasi-classical description. The analysis is performed under justified assumption that the relaxation times in two sheets of the $\pi$ band and two sheets of the $\sigma$ band are proportionally related, i.e. $\tau_{\pi 1} \propto \tau_{\pi 2}$ and $\tau_{\sigma 1} \propto \tau_{\sigma 2}$. Data on the resistivity, Hall coefficient and thermoelectric power suggest that the carbon substitution modifies the intraband scattering mainly in one of the conduction bands, which is most likely to be the $\pi$ band. On the contrary, changes of the charge carrier concentration, which are caused by electron doping, seem to be less influential in determining the transport properties of $MgB_2$ we have studied.

*Acknowledgment*

Authors are grateful to dr. C. Sułkowski for help in the thermopower measurements. This work was supported by the Polish Ministry of Science and Higher Education under research projects for the years 2006-2009 (Project No. N202 131 31/2223) and, partially, for the years 2009-2011 (Project No. N N510 357437).

**Figure captions**

1. (Color online) Normalized magnetization $M/M_0$ curves obtained at 0.5 mT field (zero-field-cooled condition) for Mg(B$_{1-x}$C$_x$)$_2$ single crystals with $x$ = 0, 0.02, 0.06, 0.1. $M_0$ is here an extrapolated value of the magnetization at $T$ = 0 K. Inset shows the resistivity at $T$ = 300K as a function of doping. Full points denote our results, while hollow points show data from Ref. [14].

2. (Color online) Temperature dependences of the Hall coefficient for Mg(B$_{1-x}$C$_x$)$_2$ single crystals with $x$ = 0, 0.02, 0.06, and 0.1.

3. (Color online) Influence of the C substitution on the Hall coefficient (right axis, empty diamonds) and the thermoelectric power (left axis, full symbols) measured in-plane (circles) and out-of-plane (triangles) for the series of the Mg(B$_{1-x}$C$_x$)$_2$ single crystals.

**Figures**

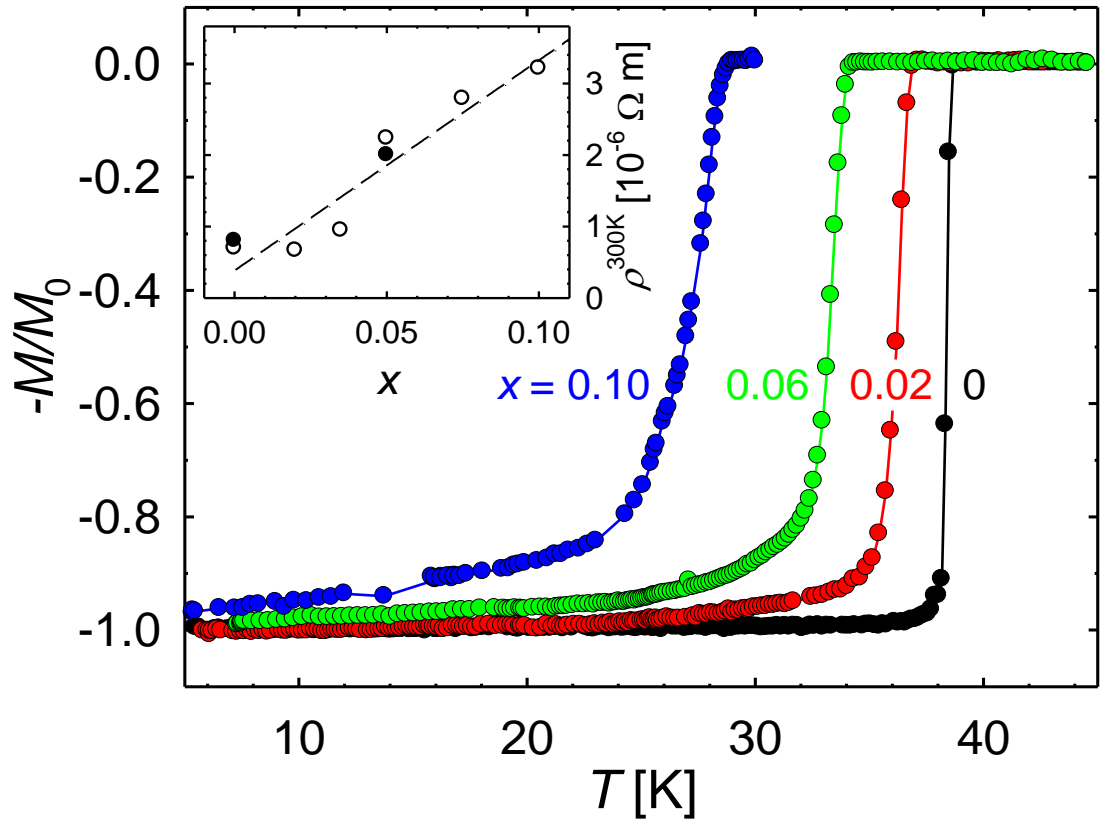

**Fig. 1.**

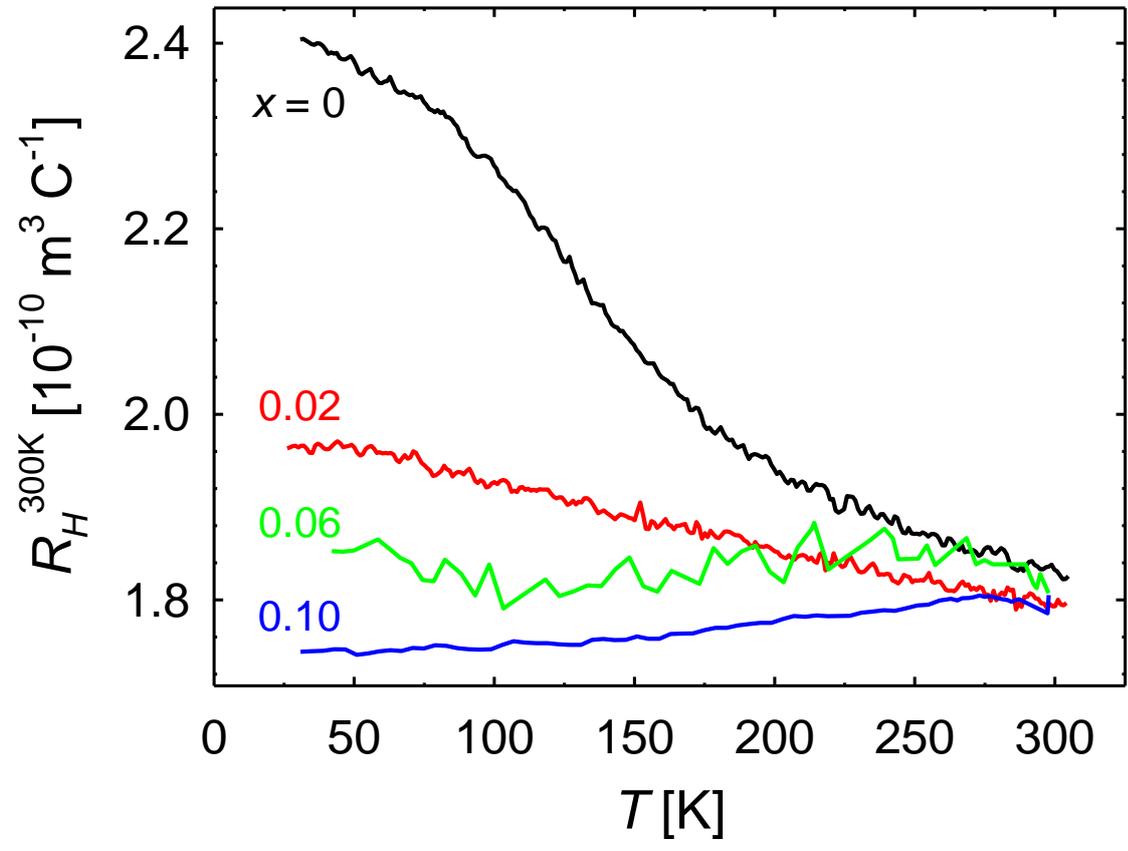

Fig. 2.

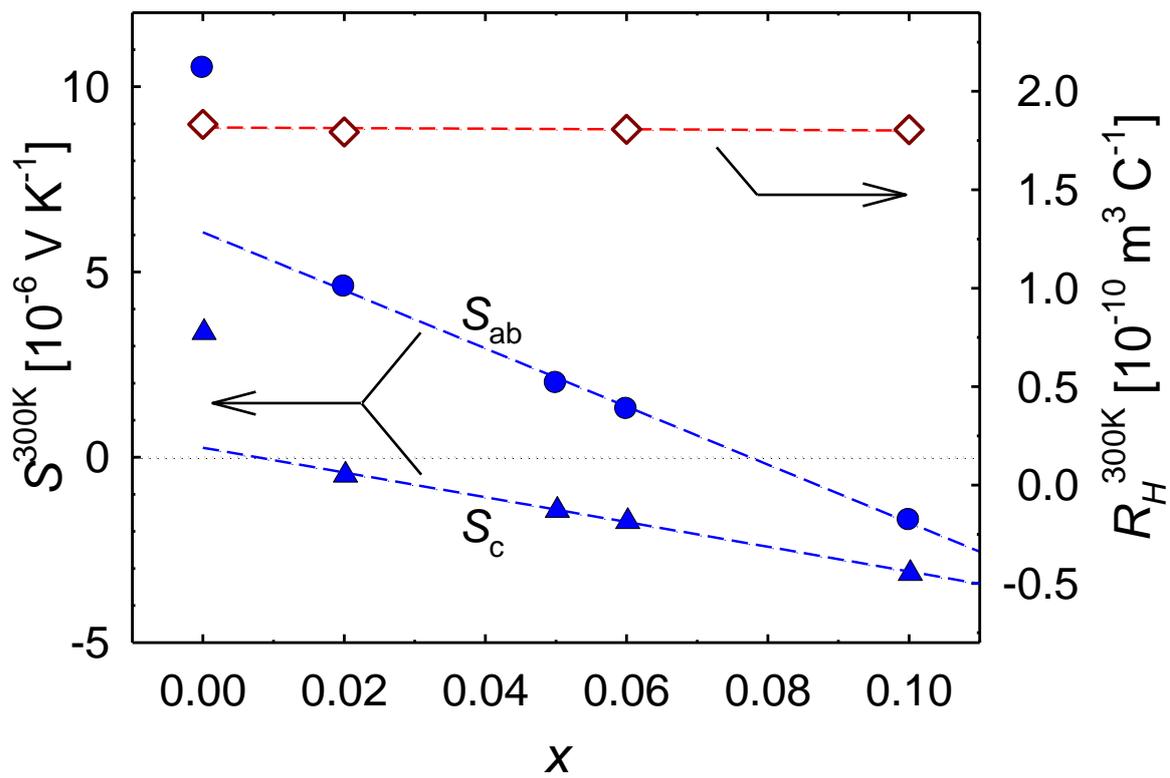

**Fig. 3.**